\documentclass[onecolumn]{autartmod}
\input epsf

\usepackage[T1]{fontenc}
\usepackage[latin9]{inputenc}
\usepackage{amsmath}
\usepackage{amssymb}
\usepackage{graphicx,color}
\usepackage{epstopdf}

\usepackage{bbm}
\renewcommand{\t}{^{\mbox{\tiny {T}}}}

\newcommand{\EQ}{\begin{eqnarray}}
\newcommand{\EN}{\end{eqnarray}}
\newcommand{\EQQ}{\begin{eqnarray*}}
\newcommand{\ENN}{\end{eqnarray*}}

\newcommand{\eproof}{\hfill\rule{2mm}{2mm}}

\newcommand{\bstate}{\medskip\begin{state} }
\newcommand{\estate}{ \hfill  \rule{1mm}{2mm}\medskip\end{state}}

\newcommand{\bass}{\medskip\begin{ass} }
\newcommand{\eass}{ \hfill  \rule{1mm}{2mm}\medskip\end{ass}}

\newcommand{\brem}{\medskip \begin{remark}  }
\newcommand{\erem}{\hfill \rule{1mm}{2mm}\medskip
\end{remark} }
\newcommand{\bthm}{\medskip\begin{theorem}  }
\newcommand{\ethm}{ \hfill  \rule{1mm}{2mm} \medskip
\end{theorem} }
\newcommand{\blem}{\medskip\begin{lemma}  }
\newcommand{\elem}{ \hfill \rule{1mm}{2mm}\medskip
\end{lemma} }
\newcommand{\bcorollary}{\medskip\begin{corollary}  }
\newcommand{\ecorollary}{  \hfill \rule{1mm}{2mm}\medskip
\end{corollary} }
\newcommand{\bdefn}{\medskip\begin{definition}}
\newcommand{\edefn}{  \hfill \rule{1mm}{2mm}\medskip
\end{definition} }
\newcommand{\bproposition}{\medskip\begin{proposition} }
\newcommand{\eproposition}{\hfill \rule{1mm}{2mm}\medskip
\end{proposition} }
\newcommand{\bexample}{\medskip\begin{example} \rm}
\newcommand{\eexample}{ \hfill \rule{1mm}{2mm}\medskip
\end{example} }
\newcommand{\proofnow}{\noindent{\bf Proof: }}

\newtheorem{theorem}{\bf Theorem}[section]
\newtheorem{ass}{\bf Assumption}[section]
\newtheorem{lemma}{\bf Lemma}[section]
\newtheorem{definition}{\bf Definition}[section]
\newtheorem{remark}{\bf Remark}[section]
\newtheorem{corollary}{\bf Corollary}[section]
\newtheorem{proposition}{\bf Proposition}[section]
\newtheorem{example}{\bf Example}[section]
\newtheorem{state}{\bf Assumption}[section]

\DeclareFontFamily{OMX}{yhex}{}
\DeclareFontShape{OMX}{yhex}{m}{n}{<->yhcmex10}{}
\DeclareSymbolFont{yhlargesymbols}{OMX}{yhex}{m}{n}
\DeclareMathAccent{\wideparen}{\mathord}{yhlargesymbols}{"F3}

\begin{document}

\begin{frontmatter}

\title{Stabilization with a Specified External Gain for Linear MIMO Systems
and Its Applications to Control of Networked Systems}

\thanks[footnoteinfo]{This research was supported under The University of Hong Kong Research Committee Post-doctoral Fellow Scheme}

\author[A]{Lijun Zhu} \ead{ljzhu@eee.hku.hk},
\author[B] {Zhiyong Chen}\ead{zhiyong.chen@newcastle.edu.au},
\author[C] {Xi Chen}\ead{chenxi 99@wust.edu.cn},
\author[A] {David J. Hill} \ead{dhill@eee.hku.hk}

\address[A]{Department of Electrical and Electronic Engineering, The University of Hong Kong, Hong Kong}
\address[B]{School of Electrical Engineering and Computing, The University of Newcastle, Callaghan, NSW 2308, Australia}  
\address[C]{Engineering Research Center of Metallurgical
Automation and Measurement Technology, Ministry of Education, Wuhan University
of Science and Technology, Wuhan 430081 China}


\begin{abstract}
This paper studies a stabilization problem for linear
MIMO systems subject to external perturbation that further requires the closed-loop system render a
specified gain from the external perturbation to the output. The problem arises from
control of networked systems, in particular,
robust output synchronization of heterogeneous linear MIMO multi-agent
systems via output feedback/communication.
We propose a new approach that converts a class of MIMO
systems into a normal form via repeated singular value decomposition
and prove that a stabilization controller
with a specified external gain can be explicitly constructed for the normal form.
Two scenarios with static state feedback and
dynamic output feedback are investigated.  By integrating
the reference model and internal model techniques, the
robust output synchronization problem for MIMO multi-agent systems
is  converted into a stabilization problem with a specified external gain
and solved by the developed approach.
\end{abstract}
\end{frontmatter}

\section{Introduction}

Stabilization for linear MIMO systems is a well developed and widely used
technique in modern control theory. The main focus of this paper is study
a stabilization problem for linear
MIMO systems subject to external perturbation that further requires the closed-loop system render a
specified gain from the external perturbation to the output.
The tool to characterize the gain from the  external perturbation to the output
is the so-called  input-to-output stability (IOS) gain with
the external perturbation regarded as an input to the closed-loop system \cite{Sontag1995,Sontag1999}.

It is known that when a linear system is stabilizable, a feedback controller that renders the
asymptotic stability for the system free of external perturbation
also makes the closed-loop system IOS when external perturbation
is taken into consideration. The IOS gain is determined by the closed-loop system
structure.  However, it remains a challenging task when  an input-to-output stabilization further requires an arbitrarily specified external gain.
That requires an input-to-output stabilization controller to further achieve an arbitrarily specified external gain.
Such a stabilization is interesting   by itself when
one is interested in managing the  external  influence to system output
through the stabilization controller design. Also, the problem is well motivated
from studying networked systems.

A relevant research topic in the literature is  $H_{\infty}$ and $H_{2}$ (sub)optimal
control. For example, Sontag discussed the close relation between
input-to-state/output formulation and $H_{\infty}$ and $H_{2}$ (sub)optimal
control in \cite{Sontag2008}.  It is known that the solution of $H_{\infty}$ and $H_{2}$
control relies on the solution of Riccati equations for  a linear system
\cite{Doyle1989} and Hamilton-Jacobi-Isaacs partial differential
equations for nonlinear systems \cite{Isidori1992,Isidori1995,VanderSchaft1992,Ball1993}.
What differs from input-to-output stabilization here is that the $H_{\infty}$
or $H_{2}$ gain through admissible state feedback and output feedback
control is greater or equal than a minimum which depends on the system
structure and the solutions to Riccati equations. In other words, the
impact of a perturbation in terms of $H_{\infty}$
or $H_{2}$ gain can be optimized/minimized by feedback control, but not made arbitrarily small.
However, the problem studied in this paper requires an arbitrarily specified external gain
through feedback controller design.

One sufficient condition for the solvability of $H_{\infty}$ or $H_{2}$ control
is that the linear system is stabilizable and detectable and the solutions to
the corresponding Riccati equations exist.
The conditions needed for the stronger requirement on arbitrarily specified external gain
are studied in paper. In particular, with some additional conditions, linear
MIMO systems can be transformed to a so-called normal form for which the
problem   can be solved.  It is noted that
the same problem has been solved for nonlinear minimum-phase systems
in the previous paper \cite{Zhu2016c} using the backstepping technique and
the improved small gain theorem. But the systems in \cite{Zhu2016c} are SISO
and in the so-called lower triangular form.
Therefore, the result cannot be applied in
the present MIMO and non-lower-triangular systems.

An important motivation or application of the proposed technique of stabilization with
a specified external gain is  the robust output synchronization problem for
heterogeneous multi-agent systems through output feedback and output communication.
Heterogeneity among agents is commonly encountered in real-world
applications such as vehicle platoons in \cite{Tanner2007} and power
systems in \cite{Bergen1981} where the dynamics of subsystems are
not uniform. Synchronization of multi-agent systems aims to achieve the
agreement on agents' outputs via local communication among agents
despite the heterogeneity. For homogeneous multi-agent systems, the
synchronization pattern is naturally embedded in the homogeneous
part of agents' dynamics, but it is not explicit for heterogeneous
systems. In \cite{Wieland2011}, a necessary condition for synchronization
of heterogeneous multi-agent systems called internal model principle
has been given as that there must exist a homogeneous kernel for each
agent that embraces the synchronization pattern. The homogeneous kernel
is either embedded in the original dynamics or explicitly constructed
through the controller design according to the task.

More specifically, as exposed in many works, e.g., \cite{Wieland2011,Su2012,Isidori2014,Zhu2016,Zhu2016a},
the synchronization controller of heterogeneous multi-agent systems
can be explicitly constructed in a two-step manner. The first step
is the consensus of reference models that are constructed to be homogeneous
for each agent and embed the homogeneous kernel. The second step is
the regulation of each individual agent's output to the output of
its own reference model. In particular, when agent dynamics contain
no uncertainties, the regulation can be achieved by the feedforward
compensation control. This method was utilized for the synchronization
of linear systems in \cite{Wieland2011} and nonlinear systems in
\cite{Chen2014}. However, the exact feedforward compensation becomes
impossible if agent dynamics have uncertainties (see detailed discussion
in \cite{Zhu2016}). In this case, researchers appeal to  robust
output regulation theory to handle uncertainties. It relies on a class
of dynamic observers called internal model (in the context of robust
output regulation theory) whose dynamics do not depend on the uncertainties
but asymptotically generate the steady-state solution for states and/or
control input. With the internal model, the regulation problem can
be converted into a stabilization problem for which a feedback controller
can be designed. This method has been used for the robust output synchronization
for linear systems in \cite{Kim2011} and nonlinear systems in \cite{Isidori2014,Zhu2016a}.

More specifically, synchronization controllers can be classified to
rely on the state or output communication. For state communication,
each agent is allowed to transmit its internal states to its neighbors,
while for output communication, it becomes slightly more restrictive and
only the output information is allowed to transmit. The freedom to
choose what internal information to transmit over the network makes synchronization
on state communication less complicated than that on output communication.
Consequently, the aforementioned two actions, namely consensus for
reference models and regulation of each agent to its reference, can
be completely separated for  state communication, which simplifies
the controller design procedure \cite{ChenXi2016}. For the output communication case,
the work in \cite{Zhu2016a} shows that these two actions are mutually
perturbed by each other (hence further called \textit{perturbed consensus}
and \textit{perturbed regulation}). In particular, the consensus action
is perturbed by the regulation error and the regulation action is
disturbed by the disagreement in consensus. The output synchronization
framework via output communication developed in \cite{Zhu2016a} suggests
that  synchronization is achieved if both perturbed consensus and
perturbed regulation are solved as well as a small gain condition
is satisfied.

The robust output synchronization of heterogeneous linear MIMO multi-agent
systems via output communication is yet to be investigated and is
the focus of the second part of this paper. In particular, we need
to propose a modified internal model and show that the robust output
synchronization problem of MIMO multi-agent systems can also be converted
into perturbed consensus and perturbed regulation problems. Moreover,
the  perturbed regulation  problem corresponds to the stabilization problem
which can be solved using techniques developed in the first part of the paper

The rest of this paper is structured as follows. In Section~\ref{sec:IOS-SG},
we will first introduce the problem of stabilization
with a specified external gain and its motivation.
The main results on stabilization
with a specified external gain are given in Section~\ref{sub:NF}.
In particular,  we propose a state feedback controller for the  MIMO systems in the normal
form as well as a dynamic output feedback controller for systems of
a special structure. In Section \ref{sec:ROS}, we formulate
the robust output synchronization problem for linear heterogeneous MIMO multi-agent systems
and introduce the framework
that converts the problem into a standard perturbed consensus problem
and a perturbed regulation problem.  Moreover,
with the aid of modified internal model design, we show that the perturbed
regulation problem is equivalent to the
stabilization problem with a specified gain as studied in Section~\ref{sec:IOS-SG}.
The numerical simulation
is conducted in Section \ref{sec:Sim} and the paper is concluded
in Section~\ref{sec:con}.

\section{Problem Formulation and Motivation \label{sec:IOS-SG}}

We  consider a class of linear MIMO control systems typically represented by the following equations
\begin{eqnarray}
\dot{x} & = & Ax+Bu+R\zeta\nonumber \\
y & = & Cx,\label{eq:system}
\end{eqnarray}
where $x\in\mathbb{R}^{n}$ is the state, $u\in\mathbb{R}^{m}$ the
input, $y\in\mathbb{R}^{p}$ the output, and $\zeta\in\mathbb{R}^{\ell}$
external perturbation.
The matrices
$A\in\mathbb{R}^{n\times n}$, $B\in\mathbb{R}^{n\times m}$,
$R\in\mathbb{R}^{n\times\ell}$ and $C\in\mathbb{R}^{p\times n}$
have compatible dimensions. The external perturbation $\zeta$ may represent
external signals or influence of other subsystem to the
system (\ref{eq:system})  in an interconnected setting.

A general linear controller takes a static form
\begin{equation}
u=Kx\label{eq:sf_sample}
\end{equation}
or a dynamic form
\begin{eqnarray}
u & = &K\chi \nonumber \\
\dot{\chi} & = & A \chi+L(y-C\chi)+Bu+R\zeta
 \label{eq:of_sample}
\end{eqnarray}
that induces a closed-loop system
\begin{eqnarray}
\dot x_c &=& A_c x_c + R_c \zeta \nonumber\\
y & = & C_c x_c\label{eq:system-cl}
\end{eqnarray}
for $x_c= x$ or $x_c =\mbox{col}(x, \chi)$, respectively.

Without the external perturbation $\zeta$, the stabilization controller design for a linear system
is well known in the literature. Moreover, for any stabilization controller that ensures a Hurwitz $A_c$,
the basic property of a linear system implies that the closed-loop system  (\ref{eq:system-cl})
is automatically bounded input bounded output (BIBO)
or input-to-output stable (IOS), with $\zeta$ as the input and $y$ the output.
Also, the gain from $\zeta$ to $y$ can be explicitly computed.
However, it remains a difficult task to design a stabilization controller such that
the closed-loop system  (\ref{eq:system-cl})   has an {\it arbitrarily specified} gain from $\zeta$ to $y$.

More specifically, we will use a quadratic function to characterize the
relationship between the input and output for the linear closed-loop system (\ref{eq:system-cl}).

\bdefn \label{def:ISS-Lyp} The linear MIMO system (\ref{eq:system-cl}) with a Hurwitz matrix $A_c$
is said to admit a quadratic IOS-Lyapunov function $V(x_c)=x_c\t Px_c$  with $P=P\t >0$
if there exist positive constants $\alpha$
and $\beta$ such that
\begin{equation}
\dot{V}(x)\leq-\alpha\|y\|^{2}+\beta\|\zeta\|^{2}.\label{eq:ISS-Lyp}
\end{equation}
In particular, $\beta/\alpha$
is called an IOS gain. \edefn

\brem
The gain between $\zeta$ to $y$ is called an IOS gain because
the external perturbation $\zeta$ is regarded as an input to the closed-loop system (\ref{eq:system-cl}).
To avoid any confusion between $\zeta$ and the actual input $u$ for the open-loop system (\ref{eq:system}),
we call the gain an {\it external gain} throughout the paper.
\erem

With the aforementioned setting, the main objective of this paper is rigorously formulated
in the following problem.

The problem of {\it stabilization with a specified external gain $\gamma$ ($\gamma$-stabilization)} for the system (\ref{eq:system})
aims, for an arbitrarily specified $\gamma>0$, to find a controller (\ref{eq:sf_sample}) or (\ref{eq:of_sample})
such that the system (\ref{eq:system-cl}) has a Hurwitz matrix $A_c$
and an external gain $\gamma$.

The $\gamma$-stabilization problem has an independent interest when
one is interested in managing the influence of an external
signal to system output through the stabilization controller design. Also, the problem is well motivated
from studying networked systems. One simple motivating example is given below. A complete application
of $\gamma$-stabilization can be found in
Section \ref{sec:ROS} in effectively solving the robust output synchronization problem
for heterogeneous MIMO multi-agent systems.

\begin{figure}[t]
\centering
\includegraphics[clip,scale=0.5]{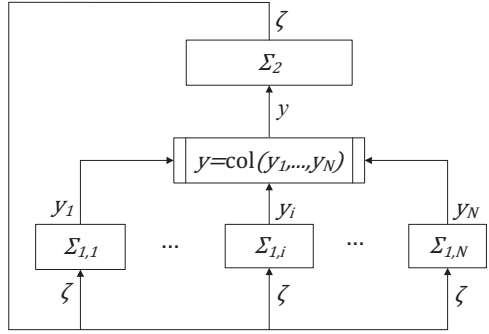}\caption{Interconnection of a network of MIMO
control systems $\Sigma_{1,i}$, $i=1,\cdots, N$ and an MIMO system $\Sigma_2$. \label{fig:inter}}
\end{figure}
\bexample \label{exp:IOS}
Consider the interconnection of two subsystems illustrated in Fig.
\ref{fig:inter}. The first subsystem $\Sigma_{1}=(\Sigma_{1,i},\cdots,\Sigma_{1,N})$
is a network of $N$ linear MIMO control systems described by (\ref{eq:system}), that is,
\begin{equation}
\Sigma_{1,i}:\begin{array}{c}
\dot{x}_{i}=A_{i}x_{i}+B_{i}u_{i}+R_{i}\zeta,\\
y_{i}=C_{i}x_{i},
\end{array},\;i=1,\cdots,N\label{eq:system_1i}
\end{equation}
where $x_{i}\in\mathbb{R}^{n_{i}}$, $u_{i}\in\mathbb{R}^{m_{i}}$
and $y_{i}\in\mathbb{R}^{p_{i}}$ are the state, input and output, respectively
and $y=\mbox{col}(y_{1},\cdots,y_{N})$ is the composite output of
$\Sigma_{1}$. The second subsystem $\Sigma_{2}$ is given by
\begin{equation}
\Sigma_{2}:\begin{array}{c}
\dot{\tau}=A_{\tau}\tau+R_{\tau}y,\\
\zeta=C_{\tau}\tau,
\end{array}\label{eq:inter_sys}
\end{equation}
where $\tau\in\mathbb{R}^{n_{\tau}}$ is the state and $\zeta\in\mathbb{R}^{\ell}$
the output.
The subsystems $\Sigma_{1}$ and $\Sigma_{2}$ are coupled through
their respective outputs $\zeta$ and $y$. Suppose $\Sigma_{2}$
admits a quadratic IOS-Lyapunov function and an external gain
$\gamma_{\zeta}$. Then, the $\gamma$-stabilization
for each $\Sigma_{1,i}$ with the external gain $\gamma <1/(N\gamma_\zeta)$ guarantees
stability of the overall system by using Proposition~\ref{Prop:IOS_SG} stated below.

For $N=1$, it is a typical interconnection of two subsystems
studied in many references such as \cite{Isidori2013,Khalil2002}.
The present scenario with $N>1$ derives from the
robust output synchronization studied in, e.g., \cite{Zhu2016a}. The
two interconnected subsystems $\Sigma_{1}$ and $\Sigma_{2}$ correspond
to the dynamics for perturbed consensus and perturbed regulation problems
to be elaborated in Section~\ref{sec:ROS}.
\eexample

The following proposition shows how $\gamma$-stabilization
for each $\Sigma_{1,i}$ guarantees the stability of the overall system in
Example~\ref{exp:IOS}.

\bproposition \label{Prop:IOS_SG}
For the control system composed of
(\ref{eq:system_1i}) and (\ref{eq:inter_sys}),
if $\Sigma_{2}$ admits a quadratic IOS-Lyapunov function and an external gain
$\gamma_{\zeta}$ and the $\gamma$-stabilization problem
for each $\Sigma_{1,i}$ with the external gain $\gamma <1/(N\gamma_\zeta)$
is solved by a linear controller. Then, the overall closed-loop linear system is stable.   \eproposition

\proofnow Let the quadratic IOS-Lyapunov function  for $\Sigma_{2}$ be $V_{\tau}(\tau)$
that satisfies
\EQ
\dot{V}_\tau(\tau) \leq-\alpha_{\zeta}\|\zeta\|^{2}+\beta_{\zeta}\|y\|^{2}
\EN
for some $\alpha_{\zeta},\beta_{\zeta}>0$ satisfying $\gamma_\zeta=\beta_\zeta/\alpha_\zeta$.
Let the state of the system composed of $\Sigma_{1,i}$ and its
$\gamma$-stabilization controller be $x_{c,i}$. Then, solvability of the $\gamma$-stabilization problem
for $\Sigma_{1,i}$ means the existence of a
quadratic IOS-Lyapunov function $V_{i}(x_{c,i})$
that satisfies
\EQ
\dot{V}_{i}(x_{c,i})\leq   -\alpha_{i} \|y_{i}\|^{2}+\beta_{i}\|\zeta\|^{2}  , \;i=1,\cdots,N
\EN
for  some $\alpha_{i}, \beta_{i}>0$ satisfying $\gamma=\beta_i/\alpha_i$.

Let $x_c=\mbox{col}(x_{c,1},\cdots,x_{c,N})$ and
\EQQ
V(\tau, x_c)={V}_\tau(\tau)/\beta_\zeta  +  \epsilon \sum_{i=1}^{N}V_{i}(x_{c,i})/\alpha_{i} ,\; \epsilon>0
\ENN
whose derivative along the trajectory of the overall closed-loop system is
\EQQ
\dot V(\tau, x_c) \leq -\alpha_{\zeta}/\beta_\zeta \|\zeta\|^{2}+ \|y\|^{2}
-  \epsilon \sum_{i=1}^{N}  \|y_{i}\|^{2}+\epsilon \sum_{i=1}^{N}  \beta_{i}/\alpha_{i}\|\zeta\|^{2}\\
\leq -1/\gamma_\zeta \|\zeta\|^{2}+ \|y\|^{2}
-  \epsilon   \|y\|^{2} +\epsilon N \gamma \|\zeta\|^{2}\\
\leq -(1/\gamma_\zeta  -\epsilon N \gamma) \|\zeta\|^{2}
-  (\epsilon -1)  \|y\|^{2} .
\ENN
As $\gamma <1/(N\gamma_\zeta)$, we can pick
\EQQ
\frac{1}{N \gamma \gamma_\zeta}  > \epsilon  >1,
\ENN
that is,
\EQQ
1/\gamma_\zeta  -\epsilon N \gamma >0,\; \epsilon -1>0.
\ENN
Applying Lasalle-Yoshizawa
Theorem (\cite{Chen2015}) leads to the state
of the closed-loop has the following asymptotical property
\EQ
\lim_{t\rightarrow\infty}\zeta(t)=0,\;\lim_{t\rightarrow\infty}y(t)=0,\label{eq:y12_t}
\EN
which further implies
\EQ \lim_{t\rightarrow\infty}\tau(t)=0,\;\lim_{t\rightarrow\infty}x_c(t)=0. \label{eq:xc12_t}
\EN
The proof is thus completed.
 \eproof


\section{Main Results: $\gamma$-Stabilization \label{sub:NF}}

The first main result is to explicitly find a state feedback controller
(\ref{eq:sf_sample}) to solve the $\gamma$-stabilization problem
for the system (\ref{eq:system}). For this purpose, we
introduce a state transformation based on
the singular value decomposition of the input matrix
that puts the system into a class of normal form.

\subsection{A Normal Form}

Let us recursively define a sequence of matrix pairs $(\Phi_j, \Gamma_j)$, $j=0,1,\cdots$ in
the following algorithm.
In the algorithm, denote the rank of $\Gamma_{j}$ by $r_j$, i.e., $\mbox{rank}(\Gamma_{j})=r_{j}.$
For convenience of notation, let $r_{-1} =m$.

\begin{itemize}
\item[(i)]
Let $j=0$ and set the initial values
\EQQ
\Phi_{0}=A,\;\Gamma_{0}=B.
\ENN

\item[(ii)] If $\Gamma_j$ has a full row rank or $\Gamma_j =0$, exit.

\item[(iii)]   One has
\EQQ
\Phi_{j} \in \mathbb{R}^{(n-\Sigma_{k=0}^{j-1}r_{k})\times (n-\Sigma_{k=0}^{j-1}r_{k})},\;
\Gamma_{j} \in \mathbb{R}^{(n-\Sigma_{k=0}^{j-1}r_{k})\times r_{j-1}}
\ENN
and the singular value decomposition (SVD) of  $\Gamma_{j}$ is
\begin{equation}
\Gamma_{j}= U_{j+1} \left[\begin{array}{cc}
\Sigma_{j} & 0\\
0 &0
\end{array}\right]H_{j+1}\t, \label{eq:svd_Bj}
\end{equation}
where
$\Sigma_{j} \in \mathbb{R}^{r_{j} \times r_{j}} $ is a diagonal matrix bearing
all non-zero singular values on the diagonal entries and
$U_{j+1} \in \mathbb{R}^{(n-\Sigma_{k=0}^{j-1}r_{k})\times(n-\Sigma_{k=0}^{j-1}r_{k}) }$
and
$H_{j+1} \in \mathbb{R}^{r_{j-1} \times r_{j-1}}$
are unitary matrices.

Since  $\Gamma_j$ does not have a full row rank and $\Gamma_j \neq 0$ (otherwise exit at step (ii)), one has
$n-\Sigma_{k=0}^{j-1}r_{k} > r_j>0$ and hence the following decomposition
\EQQ
U_{j+1} =\left[\begin{array}{cc}
\tilde{U}_{j+1} & \bar{U}_{j+1}\end{array}\right],\;
\tilde{U}_{j+1}\in\mathbb{R}^{(n-\Sigma_{k=0}^{j-1}r_{k})\times r_{j}},\;
\bar{U}_{j+1}\in\mathbb{R}^{(n-\Sigma_{k=0}^{j-1}r_{k})\times(n-\Sigma_{k=0}^{j}r_{k})}.
\ENN
Calculate
\EQ
&& \Phi_{j+1}=\bar{U}_{j+1}\t\Phi_{j}\bar{U}_{j+1} \in \mathbb{R}^{(n-\Sigma_{k=0}^{j}r_{k})\times (n-\Sigma_{k=0}^{j}r_{k})}  \nonumber\\
&& \Gamma_{j+1}=\bar{U}_{j+1}\t\Phi_{j}\tilde{U}_{j+1} \in \mathbb{R}^{(n-\Sigma_{k=0}^{j}r_{k})\times r_{j}}.
 \label{eq:ABbar}
\EN

\item[(iv)]  Let $j=j+1$ and go to step (ii).

\end{itemize}

Then, the next lemma shows a useful result from the above algorithm under a  controllability assumption.
The proof is given in the Appendix.

\bass The pair of matrices $(A,B)$ is controllable. \label{ass:con}\eass

\blem \label{lem:AB} Consider the linear system (\ref{eq:system}) satisfying Assumption~\ref{ass:con}.
Then,  there  exists a finite number $l$
such that $\Gamma_{l}$ defined in the algorithm has a full row rank. Moreover,
the pair $(\Phi_{j},\Gamma_{j})$  is controllable for $0 \leq j\leq l$. \elem

Throughout the paper, we call $l$ in Lemma~\ref{lem:AB} the {\it number of SVD steps}.
With the finite number $l$, we define the orthogonal matrix
\begin{equation}
T=\left[\begin{array}{c}
T_{1}\\
T_{2}\\
\vdots\\
T_{l}\\
T_{l+1}
\end{array}\right]:=\left[\begin{array}{c}
\bar{U}_{l}\t\cdots\bar{U}_{1}\t\\
\tilde{U}_{l}\t\bar{U}_{l-1}\t\cdots\bar{U}_{1}\t\\
\vdots\\
\tilde{U}_{2}\t\bar{U}_{1}\t\\
\tilde{U}_{1}\t
\end{array}\right]\in\mathbb{R}^{n \times n},\;
\begin{array}{c}
T_{1}\in\mathbb{R}^{ (n-\Sigma_{k=0}^{l-1}r_{k}) \times n}\\
T_{2}\in\mathbb{R}^{r_{l-1} \times n}\\
\vdots\\
T_{l}\in\mathbb{R}^{ r_1 \times n}\\
T_{l+1}\in\mathbb{R}^{r_o \times n}
\end{array}.\label{eq:T}
\end{equation}
where  $T\in\mathbb{R}^{n\times n}$.
Accordingly, we  introduce the coordinate transformation
\begin{equation}
\xi:=\left[\begin{array}{ccc}
\xi_{1}\t & \cdots & \xi_{l+1}\t\end{array}\right]\t=Tx\label{eq:ST}
\end{equation}
with the dimension of $\xi_j$ compatible with $T_j$.

We need one more assumption to obtain the input-to-output stabilization
normal form for the system  (\ref{eq:system}).

\bass The matrices $(A,B,C)$ satisfy $CA^{j-1}B=0$ for
$j=1,\cdots,l$. \label{ass:level} \eass

\blem \label{lem:y} Consider linear MIMO system (\ref{eq:system}) under
Assumptions~\ref{ass:con} and  \ref{ass:level}. Then,
$C T_j\t =0$ for $j=2,\cdots, l+1$ and hence
\begin{equation}
y=Cx=C_\xi \xi_{1}\label{eq:y_xi}.
\end{equation}
for $C_\xi =C T_1\t $.
\elem
\proofnow The proof is given in Appendix. \eproof

\medskip

Now, we can define the following matrices recursively,
\EQQ
A_{1}=\Phi_{l},\;B_{1}=\Gamma_{l},\;R_{1}=T_{1}R,
\ENN
and, for $j=2,\cdots,l+1$,
\EQQ
\begin{split}
A_{j}&=\tilde{U}_{l+2-j}\t\Phi_{l+1-j}\tilde{U}_{l+2-j} \\
B_{j}&=\left[ \begin{array}{cc}
\Sigma_{l+1-j} &0
\end{array}
\right]   H_{l+2-j}\t \\
R_{j}&=T_{j}R.
\end{split}
\ENN
Also, define
\EQQ
D_{j,k}=T_{j}AT_{k}\t,\;k=1,\cdots,j-1,\;j=1,\cdots,l+1.
\ENN

Then, the system (\ref{eq:system}) with the state transformation
(\ref{eq:ST}) under Assumptions \ref{ass:con} and \ref{ass:level}
can be converted to the following form, with $u=\xi_{l+2}$,
\begin{eqnarray}
\dot{\xi}_{j} & = & A_{j}\xi_{j}+B_{j}\xi_{j+1}+\Sigma_{k=1}^{j-1}D_{j,k}\xi_{k}+R_{j}\zeta,\;j=1,\cdots,l+1\nonumber \\
y & = & C_\xi \xi_{1}.\label{eq:sys_ST}
\end{eqnarray}

\brem
From the above development,
Lemma \ref{lem:AB} shows that $B_{1}$ is of full row rank.
Since $\Sigma_{0},\cdots,\Sigma_{l-1}$ and $H_{1},\cdots,H_{l}$
are nonsingular,  $B_{2},\cdots, B_{l+1}$ are of full
row rank. In this sense, the system (\ref{eq:sys_ST}) has a block lower-triangular
normal form.  It is easy to see that the solution to
the  $\gamma$-stabilization problem
for the system (\ref{eq:sys_ST}) implies that of (\ref{eq:system}).
\erem

\subsection{State Feedback Control \label{sub:SF}}

In this subsection, we will show how a static state controller can be designed for
the $\gamma$-stabilization problem  of the system (\ref{eq:sys_ST}) and hence
 (\ref{eq:system}).

\bthm (State Feedback) \label{thm:sf} Consider the linear MIMO system
(\ref{eq:system}) under Assumptions \ref{ass:con} and \ref{ass:level}.
Then, there exists a matrix $K$ such that the controller
\begin{equation}
u=Kx\label{eq:sf_r}
\end{equation}
 solves the $\gamma$-stabilization problem of (\ref{eq:system}). \ethm

\proofnow  For the normal form transformation given in the previous subsection,
it suffices to solve the $\gamma$-stabilization problem  of the system (\ref{eq:sys_ST}).
For this purpose, we introduce a recursive
state transformation to (\ref{eq:sys_ST}) as follows,
\begin{eqnarray}
\bar{\xi}_{1} & = & \xi_{1}\nonumber \\
\bar{\xi}_{j} & = & \xi_{j}-K_{j-1}\Xi_{j-1},\;j=2,\cdots,l+2, \label{eq:ST_bar}
\end{eqnarray}
with $\Xi_{j-1}=\mbox{col}(\bar{\xi}_{1},\cdots,\bar{\xi}_{j-1})$.

With the matrices $K_1,\cdots, K_{l+1}$ properly selected,
the system (\ref{eq:sys_ST}) with $\bar{\xi}_{l+2}=0$ can be put in the form
\begin{eqnarray*}
\dot{\bar{\xi}}_{j} & = & -\kappa_{j}\bar{\xi}_{j}+B_{j}\bar{\xi}_{j+1}+\bar{R}_{j}\zeta,\;j=1,\cdots,l\\
\dot{\bar{\xi}}_{l+1} & = & -\kappa_{l+1}\bar{\xi}_{l+1}+\bar{R}_{l+1}\zeta,\\
y & = & C_\xi \bar{\xi}_{1}
\end{eqnarray*}
for some $\bar{R}_{j}$'s. It is easy to see that,  by properly selecting
$\kappa_{i}$'s,  there exists a quadratic  Laypunov function $V(\Xi_{l+1})$ such that
\EQ
\dot{V}(\Xi_{l+1}) \leq-\alpha\|y\|^{2}+\beta\|\zeta\|^{2}-\bar{\alpha}^\prime\|\Xi_{l+1}\|^{2} \nonumber\\
\leq -\alpha\|y\|^{2}+\beta\|\zeta\|^{2}-\bar{\alpha}\|x \|^{2} \nonumber\\
\leq-\alpha\|y\|^{2}+\beta\|\zeta\|^{2} \label{eq:Vdot_sam}
\EN
for some $\bar{\alpha},\bar{\alpha}^\prime>0$ and $\beta/\alpha =\gamma$.
Therefore, the $\gamma$-stabilization problem is solved by the controller
 \EQQ
 u=\xi_{l+2}=K_{l+1}\Xi_{l+1} = K x
 \ENN
 for some matrix $K$ depending on $K_1,\cdots, K_{l+1}$.

To explicitly calculate the matrices $K_1,\cdots, K_{l+1}$, we only
consider the case with $l=1$.
The calculation can be extended for the general case with $l>1$ using the backstepping technique.
Specifically,  for $l=1$, let
\begin{eqnarray}
K_{1} & = & -B_{1}^{+}(A_{1}+\kappa_{1}I)\nonumber \\
K_{2} & = & -B_{2}^{+}\left[\begin{array}{cc}
A_{2}K_{1}+D_{2,1}+\kappa_{1}K_{1} & A_{2}-K_{1}B_{1}+\kappa_{2}I\end{array}\right].\label{eq:K12}
\end{eqnarray}
As a result, the system (\ref{eq:sys_ST}) with  $\bar{\xi}_3=0$ becomes
\begin{eqnarray*}
\dot{\bar{\xi}}_{1} & = & -\kappa_{1}\bar{\xi}_{1}+B_{1}\bar{\xi}_{2}+\bar{R}_{1}\zeta,\\
\dot{\bar{\xi}}_{2} & = & -\kappa_{2}\bar{\xi}_{2}+\bar{R}_{2}\zeta,\\
y & = & C_\xi \bar{\xi}_{1},
\end{eqnarray*}
for $\bar{R}_{1}=R_{1}$ and $\bar{R}_{2}=R_{2}-K_{1}R_{1}$.
We choose $\kappa_{2}>\|B_{1}\|^{2}+\|\bar{R}_{2}\|^{2}$ and $\kappa_{1}>\frac{1}{4}+\|\bar{R}_{1}\|^{2}+\frac{\|C_\xi\|}{2\gamma}$. Let $V(\Xi_2)=1/2(\bar{\xi}_{1}\t\bar{\xi}_{1}+\bar{\xi}_{2}\t\bar{\xi}_{2})$ be the
quadratic IOS-Lyapunov function.
Using the fact
\EQQ
\|\bar{\xi}_{1}\|\geq\|C_\xi\bar{\xi}_{1}\|/\|C_\xi\|=\|y\|/\|C_\xi\|,
\ENN
a simple calculation shows
\begin{eqnarray*}
\dot{V} (\Xi_2)& = & -\kappa_{1}\|\bar{\xi}_{1}\|^{2}-\kappa_{2}\|\bar{\xi}_{2}\|^{2}+\bar{\xi}_{1}\t(B_{1}\bar{\xi}_{2}+\bar{R}_{1}\zeta)+\bar{\xi}_{2}\t\bar{R}_{2}\zeta\\
 & \leq & -\frac{1}{2\gamma}\|y\|^{2}+\frac{1}{2}\|\zeta\|^{2}-\bar{\alpha}\|\Xi_{2}\|^{2}
\end{eqnarray*}
for some $\bar{\alpha}>0$. Thus, the $\gamma$-stabilization problem is solved. \eproof

\subsection{Extension to a Class of Output Feedback Control Systems\label{sub:OF}}

In this subsection, we will extend the method in the previous subsection
to construct the output feedback controller
to solve the $\gamma$-stabilization problem for a particular linear MIMO system described
as follows
\begin{eqnarray}
\left[\begin{array}{c}
\dot{x}\\
\dot{z}
\end{array}\right] & = & \underbrace{\left[\begin{array}{cc}
A & BQ\\
0 & M+NQ
\end{array}\right]}_{\bar{A}}\left[\begin{array}{c}
x\\
z
\end{array}\right]+\underbrace{\left[\begin{array}{c}
B\\
N
\end{array}\right]}_{\bar{B}}u+\underbrace{\left[\begin{array}{c}
R\\
NB^{+}R
\end{array}\right]}_{\bar{R}}\zeta\nonumber \\
y & = & \underbrace{\left[\begin{array}{cc}
C & 0\end{array}\right]}_{\bar{C}}\left[\begin{array}{c}
x\\
z
\end{array}\right],\label{eq:system_var}
\end{eqnarray}
where $z\in\mathbb{R}^{n_{z}}$ and $\mbox{col}(x,z)$ are the augmented
state. The matrix  $B$ has a full column rank with
 $B^+$ being the pseudo-inverse of $B$, i.e., $B^+ B = I$.
This specific linear MIMO system structure arises from the robust
output synchronization problem to be studied in Section \ref{sec:ROS}.
The problem is studied under the following additional assumption.

\bass The pair of matrices $(A,C)$ is detectable. \label{ass:det}\eass

\bthm (Dynamic Output Feedback) \label{thm:of} Consider the linear MIMO
system (\ref{eq:system_var}) under Assumptions \ref{ass:con}, \ref{ass:level}
and \ref{ass:det}. Assume $M$ is Hurwitz and $(M,N)$ is controllable.
Let the $\gamma$-stabilization of the system (\ref{eq:system})
of the same $(A,B,C,R)$ be solved by a state feedback
controller $Kx$.  Then, the $\gamma$-stabilization of the system
 (\ref{eq:system_var})  is solved by the following
output feedback controller
\begin{eqnarray}
u & = &\bar K \chi   \nonumber \\
\dot{\chi} & = & \bar{A}\chi+L(y-\bar{C}\chi)+\bar{B}u+\bar{R}\zeta
\label{eq:of}
\end{eqnarray}
where $\bar K = \left[\begin{array}{cc}
K & -Q\end{array}\right]$ and $L$ is selected
such that $\bar{A}-L\bar{C}$ is Hurwitz. \ethm

\proofnow Since $(A,C)$ is detectable and $(M,N)$ is controllable,
one has that $(\bar{A},\bar{C})$ is also detectable (see Theorem~6.23 in \cite{Huang2004}).   The state transformation $\phi=z-NB^{+}x$
and $\bar{\chi}=\chi-\mbox{col}(x,z)$
lead to
\begin{eqnarray*}
\dot{x} & = & Ax+Bu+\left[\begin{array}{cc}
BK & -BQ\end{array}\right]\bar{\chi}+R\zeta\\
\dot{\phi} & = & M\phi+(MNB^{+}-NB^{+}A)x\\
\dot{\bar{\chi}} & = & (\bar{A}-L\bar{C})\bar{\chi}\\
y & = & Cx.
\end{eqnarray*}
For the $x$-dynamics, we can regard $\bar{\chi}$ and $\zeta$ as the
external perturbation.  Since the state feedback controller $Kx$ solves
the $\gamma$-stabilization problem of the linear MIMO system described by matrices
$(A,B,C,R)$, according
to the proof of Theorem~\ref{thm:sf} (see (\ref{eq:Vdot_sam})),
there exists a quadratic IOS-Lyapunov function $V_{x}(x)$, whose derivative satisfies
\begin{equation}
\dot{V}_{x}(x)\leq-\alpha\|y\|^{2}+\beta_{1}\|\zeta\|^{2}+\beta_{2}\|\bar{\chi}\|^2-\bar{\alpha}\|x\|^2.\label{eq:Vj-3}
\end{equation}
for some $\bar{\alpha},\beta_{2}>0$ and $\beta_{1}/\alpha=\gamma$.

Since $M$ and $\bar{A}-L\bar{C}$ are Hurwitz, there exists  positive
definite matrices $P_{\phi}$ and $P_{\chi}$ such that $P_{\phi}M+M\t P_{\phi}=-I$
and $P_{\chi}(\bar{A}-L\bar{C})+(\bar{A}-L\bar{C})\t P_{\chi}=-I$.
Let $V_{\phi}(\phi)=\phi\t P_{\phi}\phi$ and $V_{\chi}(\bar\chi)=\bar\chi\t P_{\chi}\bar\chi$.
Then, the derivative of $V_{\phi}(\phi)$ along the $\phi$-dynamics is
\[
\dot{V}_{\phi}(\phi)\leq-\frac{1}{2}\|\phi\|^{2}+\beta_{\phi}\|x\|^{2}
\]
for some $\beta_{\phi}>0$. The derivative of $V_{\bar\chi}(\chi)$ along the
$\bar{\chi}$-dynamics is $\dot{V}_{\chi}(\bar\chi)=-\|\bar{\chi}\|^{2}$. Let
$V(x,\phi,\bar\chi)=V_{x}(x)+\alpha_{1}V_{\phi}(\phi)+\alpha_{2}V_{\chi}(\bar\chi)$ where $0<\alpha_{1}<\bar\alpha/\beta_{\phi}$
and $\alpha_{2}>\beta_{2}$. Then,
\EQQ
\dot{V}\leq-\alpha\|y\|^{2}+\beta_{1}\|\zeta\|^{2}
 -(\bar{\alpha}-\alpha_1\beta_{\phi})  \|x\|^2-\frac{\alpha_1}{2}\|\phi\|^{2}
-(\alpha_2 -\beta_2) \|\bar{\chi}\|^{2} \nonumber\\
\leq-\alpha\|y\|^{2}+\beta_{1}\|\zeta\|^{2}.
\ENN
Thus, the $\gamma$-stabilization of the system
 (\ref{eq:system_var})  is solved with  $\beta_{1}/\alpha=\gamma$. \eproof

\section{Robust Output Synchronization of MASs \label{sec:ROS}}

In this section, we will apply the controller design method developed
in Section~\ref{sub:NF} to solve the robust output synchronization
problem for a class of linear MIMO uncertain heterogeneous MASs
whose dynamics are described by
\begin{eqnarray}
\dot{x}_{i} & = & A_{i}(w_{i})x_{i}+B_{i}(w_{i})u_{i},\nonumber \\
y_{i} & = & C_{i}(w_{i})x_{i},\;i=1,\cdots,N\label{eq:agent}
\end{eqnarray}
where $x_{i}\in\mathbb{R}^{n_{i}}$, $u_{i}\in\mathbb{R}^{m_{i}}$,
$y_{i}\in\mathbb{R}^{p}$ are state, input and output of agent $i$,
respectively and $w_{i}\in\mathbb{R}^{\ell_{i}}$ an uncertain parameter
vector. Without loss of generality, we assume $w_{i}=0$ as the nominal
value of $w_{i}$ and
$B_{i}$ has a full column rank. The output synchronization problem is to find a
distributed control strategy $u_{i}$ for each agent (\ref{eq:agent})
such that outputs of all agents synchronize to an agreed trajectory,
regardless of the uncertainties $w_{i}$.

The group of MAS (\ref{eq:agent}) is
said to achieve {\it robust output synchronization} if there exist neighborhoods
$\mathbb{W}_{i}$ of $w_{i}=0$, $i=1,\cdots,N$, on which the output trajectories
of all agents satisfy
\[
\lim_{t\rightarrow\infty}(y_{i}(t)-y_{j}(t))=0,\;\forall i,j=1,\cdots,N.
\]

The synchronization pattern for the output trajectories, denoted by $y_{o}(t)$, is typically governed by the dynamics \begin{align}
\dot{v}_{o} & =A_{o}v_{o}\nonumber \\
y_{o} & =C_{o}v_{o}\label{eq:pattern}
\end{align}
where $v_{o}\in\mathbb{R}^{l}$, $y_{o}\in\mathbb{R}^{p}$, and $(A_{o},C_{o})$
are two prescribed matrices. The autonomous system (\ref{eq:pattern})
represents a general class of patterns including constant and/or harmonic
series up to a certain order. The definition of output synchronization,
as introduced in, e.g.,  \cite{Zhu2016a}, is revisited as follows.

In this paper, we assume that each agent $i$ can only receive relative
outputs from its neighbors with the specified weights, denoted by
\begin{equation}
\varsigma_{i}=\sum_{j\in\mathcal{N}_{i}}a_{ij}(y_{j}-y_{i}),\label{eq:chi}
\end{equation}
where $a_{ij}$ is the weight on the information transmitted between
agent $i$ and $j$ and $\mathcal{N}_{i}$ is the set of neighboring
agents from which agent $i$ can receive the relative output $y_{j}-y_{i}$.
We consider a directed graph $\mathcal{G}=(\mathcal{V},\mathcal{E},\mathcal{A})$
to represent the communication topology where the set of nodes $\mathcal{V}=\{1,\cdots,N\}$
denotes agents and the set of edges $\mathcal{E}\subseteq\mathcal{V}\times\mathcal{V}$
represents the information flow. The weighted adjacency matrix of
a graph $\mathcal{G}$ is $\mathcal{A}=[a_{ij}]$ with $a_{ii}=0$
and $a_{ij}\geq0$, more specifically, $a_{ij}>0$ for $(i,j)\in\mathcal{E}$
and $a_{ij}=0$ for $(i,j)\notin\mathcal{E}$. Denote $\mathcal{L}=[l_{ij}]$
as the Laplacian of the graph, where $l_{ii}=\sum_{j=1}^{N}a_{ij}$
and $l_{ij}=-a_{ij}$. Throughout the section, the graph is assumed to have a spanning tree,
that is, there exists a node to which all other nodes can be linked
via a directed path.

For each agent $i$, a distributed controller that uses the {\it relative
output} network communication
$\varsigma_{i}$ and the {\it output}  $y_i$  is  designed such  that
 the group of closed-loop agents achieves the robust output synchronization.
Denote  $A_{i}=A_{i}(0)$ $B_{i}=B_{i}(0)$
and $C_{i}=C_{i}(0)$ when no confusion is caused. Some assumptions
are needed for the problem.

\bass \label{Ass:detect} The pair $(A_{o},C_{o})$ is detectable.
\eass

\bass For $i=1,\cdots, N$,  the pair $(A_{i},B_{i})$ is controllable,
the pair $(A_{i},C_{i})$ is detectable,
and $C_{i}A_{i}^{j-1}B_{i}=0$ for $j=1,\cdots,l_{i}$ where $l_{i}$
is the number of SVD steps of $(A_{i},B_{i})$. \label{ass:con-1}\eass

\bass \label{Ass:TZ} For any $w_{i}\in\mathbb{W}_{i}$,
\EQQ
\mbox{rank}\left[\begin{array}{cc}
A_{i}(w_{i})-\lambda I & B_{i}(w_{i})\\
C_{i}(w_{i}) & 0
\end{array}\right]=n_{i}+m_{i},\; \forall\lambda\in\sigma(A_{o}),
\ENN
 where $\sigma(A_{o})$ denotes the spectrum of $A_{o}$. \eass

\brem Assumption \ref{Ass:detect} is common for output synchronization
of linear homogeneous multi-agent systems (see \cite{Kim2011,Seo2009}).
As shown in \cite{Zhu2016a}, it ensures that perturbed consensus
of the reference models can be achieved. Assumption~\ref{ass:con-1}
follows the assumptions in the previous section as $\gamma$-stabilization
will be shown as a required step in the present output synchronization
problem.
By Theorem 1.9 in \cite{Huang2004}, Assumption
\ref{Ass:TZ} is required such that there exists a unique solution
pair $(X_{i}(w_{i}),U_{i}(w_{i}))$ for the following regulator equations,
for all $w_{i}\in\mathbb{W}_{i}$,
\begin{align}
X_{i}(w_{i})A_{o} & =A_{i}(w_{i})X_{i}(w_{i})+B_{i}(w_{i})U_{i}(w_{i})\nonumber \\
C_{o} & =C_{i}(w_{i})X_{i}(w_{i}).\label{eq:RE}
\end{align}
The pair $(X_{i}(w_{i}),U_{i}(w_{i}))$  can be used to define the steady-state state and input to (\ref{eq:agent}) as
the the output synchronization is achieved in the pattern (\ref{eq:pattern}).
\erem

As introduced in \cite{Zhu2016a}, the robust output synchronization
problem of uncertain multi-agent systems can be solved in a framework
by addressing two coupled problems, namely perturbed consensus and
perturbed regulations problems. The perturbed consensus problem is
standard and it has been solved by introducing the homogeneous
reference model for each agent as follows
\begin{align}
\dot{v}_{i} & =A_{o}v_{i}+B_{o}C_{ \zeta} \zeta_{i},\nonumber \\
\dot{ \zeta}_{i} & =A_{ \zeta} \zeta_{i}+B_{ \zeta}\varsigma_{i},\nonumber \\
\hat{y}_{i} & =C_{o}v_{i},\;i=1,\cdots,N\label{eq:RM0}
\end{align}
Define $e_{i}=y_{i}-\hat{y}_{i}$ for $i=1,\cdots,N$ as the local
regulation error. Let $e=\mbox{col}(e_{1},\cdots,e_{N})$
and $\zeta=\mbox{col}(\zeta_{1},\cdots,\zeta_{N})$.
The reference model (\ref{eq:RM}) can be rewritten as
\begin{align}
\dot{v}_{i} & =A_{o}v_{i}+B_{o}C_{ \zeta} \zeta_{i},\nonumber \\
\dot{ \zeta}_{i} & =A_{ \zeta} \zeta_{i}+B_{ \zeta}C_o \sum_{j\in\mathcal{N}_{i}}a_{ij}(v_{j}-v_{i})
+B_{ \zeta}\sum_{j\in\mathcal{N}_{i}}a_{ij}(e_{j}-e_{i}) ,\;i=1,\cdots,N\label{eq:RM}
\end{align}
It was shown that the consensus of reference models is perturbed by the regulation
error $e$. By properly selecting matrices $B_{o}$, $A_{ \zeta}$, $B_{ \zeta}$
and $C_{ \zeta}$, in particular, with a Hurwitz $A_\zeta$,
such that consensus  of (\ref{eq:RM}) with $e=0$ is achieved.
Let $\varpi$ be the full state  that represents disagreements among reference models.
In particular,  there exists a quadratic IOS-Lyapunov function $V_{\varpi}(\varpi)$
such that
\EQQ
\dot{V}_{\varpi}(\varpi) \leq-\alpha_{\varpi}\|\varpi\|^{2}+\beta_{\varpi}\|e\|^{2}
\ENN
for some positive constant $\alpha_{\varpi}$ and $\beta_{\varpi}$.
As  $v_j-v_i$ be a part of $\varpi$ and $A_\zeta$ is Hurwitz,
there exists a quadratic IOS-Lyapunov function $V_{\tau}(\tau)$ with $\tau=\mbox{col}(\varpi, \zeta)$
such that
\begin{equation}
\dot{V}_{\tau}(\tau) \leq-\alpha_{\zeta}\|\zeta\|^{2}+\beta_{\zeta}\|e\|^{2}\label{eq:V_zeta}
\end{equation}
for some positive constant $\alpha_{\zeta}$ and $\beta_{\zeta}$. Denote $\gamma_\zeta = \beta_\zeta/\alpha_\zeta$.

The next so-called perturbed regulation problem aims to design
$u_i$ such that $\lim_{t \rightarrow\infty}e_i(t) =0$ for  the MIMO agent dynamics repeated as follows
\begin{align}
\dot{x}_{i} & =A_{i}(w_{i})x_{i}+B_{i}(w_{i})u_{i}\nonumber \\
e_{i} & =C_{i}(w_{i})x_{i}-C_{o}v_{i},\;i=1,\cdots N.\label{eq:agent-1}
\end{align}
Note that $v_i$ is governed by the first equation of (\ref{eq:RM}), i.e.,
\EQ
\dot{v}_{i}  =A_{o}v_{i}+B_{o}C_{ \zeta} \zeta_{i}, \;i=1,\cdots N \label{exo}
\EN
that is regarded as an exosystem with perturbation $\zeta_i$.

In what follows, we will convert the perturbed regulation problems into a $\gamma$-stabilization
problem by exploiting the robust output regulation theory
\cite{Huang2004,Chen2015}. Let
\EQQ
p(\lambda)=\lambda^{s}+\alpha_{1}\lambda^{(s-1)}+\cdots+\alpha_{(s-1)}\lambda+\alpha_{s}
\ENN
be the minimal polynomial of $A_{o},$ and
\[
\bar A_o=\left[\begin{array}{cccc}
0 & 1 & \cdots & 0\\
0 & 0 & \cdots & 0\\
\vdots & \vdots & \vdots & \vdots\\
0 & 0 & \cdots & 1\\
-\alpha_{s} & -\alpha_{(s-1)} & \cdots & -\alpha_{1}
\end{array}\right],\;\bar C_o\t=\left[\begin{array}{c}
1\\
0\\
\vdots\\
0\\
0
\end{array}\right].
\]
Let $\Upsilon_{i}(w_{i})=\mbox{col}(\Omega_{1}(w_{i}),\cdots,\Omega_{m_{i}}(w_{i}))$
where $\Omega_{j}(w_{i})=\mbox{col}(U_{i,j}(w_i),U_{i,j}(w_i)A_{o},\cdots,U_{i,j}(w_i)A_{o}^{s-1})$
and $U_{i,j}(w_i)$ is the $j$th row of $U_{i}(w_{i})$. It can be verified
that
\EQQ
\Upsilon_{i}(w_{i})A_{o}=\Phi_{i} \Upsilon_{i}(w),\;
U_{i}(w_{i})=\Psi_{i} \Upsilon_{i}(w)
\ENN
for $\Phi_{i}=I_{m_{i}}\otimes\bar A_o$, $\Psi_{i}=I_{m_{i}}\otimes\bar C_o$.
Denote $\theta_{i}(v_{i},w_{i})=T_{i}\Upsilon_{i}(w_{i})v_{i}$ for any nonsingular matrix
$T_{i}$ to be specified later.  Then, along the trajectory of (\ref{exo}),
\begin{align}
\dot{\theta}_{i}(v_{i},w_{i})&=T_{i}\Phi_{i}T_{i}^{-1}\theta_{i}(v_{i},w_{i})+T_{i}\Upsilon_{i}(w_{i})B_{o}C_{ \zeta} \zeta_{i}\nonumber \\
U_{i}(w_{i})v_{i} &=\Psi_{i}T_{i}^{-1}\theta_{i}(v_{i},w_{i}).\label{eq:SSG}
\end{align}
The dynamics (\ref{eq:SSG}) can be called a steady-state generator for the steady-state
input $U_{i}(w_{i})v_{i}$. Based on this steady-state generator, we modify the classic
internal model in \cite{Chen2015,Huang2004} as follows
\begin{equation}
\dot{\eta}_{i}=M_{i}\eta_{i}+N_{i}u_{i}+T_{i}\Upsilon_{i}B_{o}C_{ \zeta} \zeta_{i}-N_{i}B_{i}^{+}X_{i}B_{o}C_{ \zeta} \zeta_{i}\label{eq:IM}
\end{equation}
where  $\Upsilon_{i}=\Upsilon_{i}(0)$, $X_{i}=X_{i}(0)$ and
$B_{i}^{+}$ is the pseudo-inverse of $B_{i}$.
The matrices $M_i$ and $N_i$ are selected such that
$M_i$ is Hurwitz, $(M_{i},N_{i})$ is controllable, and the spectrum
of $M_{i}$ and $\Phi_{i}$ are disjoint. The matrix $T_{i}$ in (\ref{eq:SSG})
is the unique solution to the Sylvester equation
\EQ \label{syl}
T_{i}\Phi_{i}-M_{i}T_{i}=N_{i}\Psi_{i}.
\EN

We then attach the internal model (\ref{eq:IM}) to the system (\ref{eq:agent-1})
and perform the following coordinate and input transformation
\begin{align}
\bar{x}_{i} & =x_{i}-X_i(w_i) v_{i} \nonumber \\
\bar{\eta}_{i} & =\eta_{i}-\theta_{i}(v_{i},w_{i})\nonumber \\
\bar{u}_{i} & =u_{i}-\Psi_{i}T_{i}^{-1}\eta_{i}. \label{eq:CT}
\end{align}
As a result, the system (\ref{eq:agent-1}) can be put in the following form
\EQQ
\dot{\bar{x}}_{i} &= & A_{i}(w_{i})x_{i}+B_{i}(w_{i})u_{i}-X_{i}(w_{i})A_{i}(w_{i})v_{i}-X_{i}(w_{i})B_{o}C_{ \zeta} \zeta_{i}\nonumber \\
&= & A_{i}(w_{i})\bar{x}_{i}+B_{i}(w_{i})\bar{u}_{i}+B_{i}(w_{i})\Psi_{i}T_{i}^{-1}\bar{\eta}_{i}+B_{i}(w_{i})\Psi_{i}\nonumber \\
 && \times T_{i}^{-1}\theta(v_{1},w_{i})-B_{i}(w_{i})U_{i}(w_{i})v_{i}-X_{i}(w_{i})B_{o}C_{ \zeta} \zeta_{i}\nonumber \\
&= & A_{i}(w_{i})\bar{x}_{i}+B_{i}(w_{i})\bar{u}_{i}+B_{i}(w_{i})\Psi_{i}T_{i}^{-1}\bar{\eta}_{i}-X_{i}(w_{i})B_{o}C_{ \zeta} \zeta_{i} \nonumber\\
e_{i}  &=&C_{i}(w_{i})x_{i}-C_{o}v_{i} =
C_{i}(w_{i})\bar x_{i} + C_i(w_i) X_i(w_i) v_i-C_{o}v_{i} \nonumber \\
&= &C_{i}(w_{i})\bar x_{i}
\ENN
by using (\ref{eq:RE}) and (\ref{eq:SSG}) in the calculation.  Also, the internal model dynamics
(\ref{eq:IM}) becomes
\begin{eqnarray*}
\dot{\bar{\eta}}_{i} & = & M_{i}\eta_{i}+N_{i}u_{i}-T_{i}\Phi_{i}T_{i}^{-1}\theta_{i}(v_{o},w_{i})+T_{i}(\Upsilon_{i}-\Upsilon_{i}(w_{i}))B_{o}C_{ \zeta} \zeta_{i}-N_{i}B_{i}^{+}X_{i}B_{o}C_{ \zeta} \zeta_{i}\\
 & = & (M_{i}+N_{i}\Psi_{i}T_{i}^{-1})\bar{\eta}_{i}+N_{i}\bar{u}_{i}+T_{i}(\Upsilon_{i}-\Upsilon_{i}(w_{i}))B_{o}C_{ \zeta} \zeta_{i}-N_{i}B_{i}^{+}X_{i}B_{o}C_{ \zeta} \zeta_{i}.
\end{eqnarray*}
Let $\mathcal{X}_{i}:=\mbox{col}(\bar{x}_{i},\bar{\eta}_{i})$. In other words, the system composed of
(\ref{eq:agent-1}) and (\ref{eq:IM}) can be put into a more compact form
\begin{align}
\dot{\mathcal{X}}_{i} & =\bar A_{i}\mathcal{X}_{i}+\bar B_{i}\bar{u}_{i}+\bar R_{i} \zeta_{i}
+\Delta_{i}(w_{i}) \mbox{col}(\mathcal{X}_{i}, \bar u_i, \zeta_i ) \nonumber \\
e_{i} & =\bar C_{i}\mathcal{X}_{i}+\bar{\Delta}_{i}(w_{i}) \mathcal{X}_{i} \label{eq:Chi}
\end{align}
 where
\begin{gather*}
\bar A_{i}=\left[\begin{array}{cc}
A_{i} & B_{i}\Psi_{i}T_{i}^{-1}\\
0 & M_{i}+N_{i}\Psi_{i}T_{i}^{-1}
\end{array}\right],\;\bar B_{i}=\left[\begin{array}{c}
B_{i}\\
N_{i}
\end{array}\right],\\
\bar R_{i}=\left[\begin{array}{c}
-X_{i}(w_{i})B_{o}C_{ \zeta}\\
-N_{i}B_{i}^{+}X_{i}B_{o}C_{ \zeta}
\end{array}\right],\;\bar C_{i}=\left[ \begin{array}{cc}C_{i} & 0\end{array}\right],
\end{gather*}
and
\EQQ
\Delta_{i}(w_{i}) = \\
\left[\begin{array}{cccc}
A_{i}(w_{i})-A_{i}(0) & (B_{i}(w_{i})-B_{i}(0))\Psi_{i}T_{i}^{-1}\Psi_{i}T_{i}^{-1} &
 B_{i}(w_{i})-B_{i}(0) & -(X_{i}(w_{i})-X_{i}(0))B_{o}C_{ \zeta}  \\
0& 0& 0& T_{i}(\Upsilon_{i}(0)-\Upsilon_{i}(w_{i}))B_{o}C_{ \zeta}
\end{array}\right] \\
\bar{\Delta}_{i}(w_{i})  = \left[\begin{array}{cc}  C_{i}(w_{i})-C_{i}(0)  & 0 \end{array}\right].
\end{eqnarray*}
Obviously, one has  ${\Delta}_{i}(0)=0$
and $\bar{\Delta}_{i}(0) =0$.

It is noted that the system (\ref{eq:Chi}) with ${\Delta}_{i}(w_{i})=0$
and $\bar{\Delta}_{i}(w_{i}) =0$ takes the form (\ref{eq:system_var}).
As $\Delta_{i}(w_{i})$ and $\bar{\Delta}_{i}(w_{i})$ continuously depend on $w_{i}$,
we have the following result by directly applying Theorem~\ref{thm:of}.

\blem \label{lem:mas:stb} Consider the linear MIMO
system  (\ref{eq:Chi}) under Assumption~\ref{ass:con-1}. Assume $M_i$ is Hurwitz and $(M_i,N_i)$ is controllable.
There exist neighborhoods
$\mathbb{W}_{i}$ of $w_{i}=0$ such that, for all $w_i \in \mathbb{W}_{i}$,
the $\gamma$-stabilization of the system (\ref{eq:Chi}) is solved by an output feedback controller of the form
\begin{eqnarray}
\bar u_i & = & \left[\begin{array}{cc}
K_i & -\Psi T_i^{-1} \end{array}\right]\chi_i   \nonumber \\
\dot{\chi}_i & = & \bar{A}_i\chi_i+L_i (e_i -\bar{C}_i \chi_i)+\bar{B}_i \bar u_i +\bar{R}_i\zeta_i.
\label{eq:mas-stb}
\end{eqnarray}
  \elem

From the above development, the solution to the robust output synchronization problem is summarized as follows.

\bthm \label{thm:outputSyn}Consider agent (\ref{eq:agent}) under
Assumptions \ref{Ass:detect},  \ref{ass:con-1}, and \ref{Ass:TZ}. There exist neighbourhoods
$\mathbb{W}_{i}$ of $w_{i}=0$, $i=1,\cdots,N$, on which the robust output synchronisation problem
is solved by a distributed controller of the form
\begin{eqnarray}
 u_i & = & \left[\begin{array}{cc}
K_i & -\Psi T_i^{-1} \end{array}\right]\chi_i  +\Psi_{i}T_{i}^{-1}\eta_{i} \nonumber \\
\dot{\chi}_i & = & \bar{A}_i\chi_i+L_i (y_{i}-C_{o}v_{i} -\bar{C}_i \chi_i)+\bar{B}_i \left[\begin{array}{cc}
K_i & -\Psi T_i^{-1} \end{array}\right]\chi_i   +\bar{R}_i\zeta_i \nonumber \\
\dot{v}_{i} & = &A_{o}v_{i}+B_{o}C_{ \zeta} \zeta_{i},\nonumber \\
\dot{ \zeta}_{i} & =&A_{ \zeta} \zeta_{i}+B_{ \zeta}\sum_{j\in\mathcal{N}_{i}}a_{ij}(y_{j}-y_{i}) ,\nonumber \\
\dot{\eta}_{i}&=& M_{i}\eta_{i}+N_{i}u_{i}+T_{i}\Upsilon_{i}B_{o}C_{ \zeta} \zeta_{i}-N_{i}B_{i}^{+}X_{i}B_{o}C_{ \zeta} \zeta_{i} . \label{mas-overallcontroller}
\end{eqnarray} \ethm

\proofnow  By Lemma~\ref{lem:mas:stb}, the $\gamma$-stabilization of the system  (\ref{eq:Chi})
can be solved by a controller of the form (\ref{eq:mas-stb}) for $\gamma<1/(N\gamma_\zeta)$
where $\gamma_\zeta$ is given in  (\ref{eq:V_zeta}). By Proposition~\ref{Prop:IOS_SG},
the closed-loop system composed of the $\tau=\mbox{col}(\varpi, \zeta)$-dynamics and  (\ref{eq:Chi}) is stable
because the $\tau$-dynamics admit a quadratic IOS-Lyapunov function and an external gain
$\gamma_{\zeta}$ and the $\gamma$-stabilization problem
for  (\ref{eq:Chi})  with $\gamma <1/(N\gamma_\zeta)$
is solved by a linear controller.   The overall controller consisting 
of the reference model (\ref{eq:RM0}), the internal model (\ref{eq:IM}), and the $\gamma$-stabilization controller (\ref{eq:mas-stb}) becomes (\ref{mas-overallcontroller}). The proof is thus completed.
\eproof

\section{Numerical Simulation \label{sec:Sim}}

\begin{figure}
\centering\includegraphics[scale=0.8]{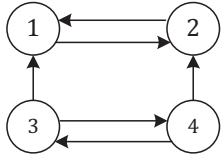}\protect\caption{Network Topology of the multi-agent systems.\label{fig:net}}
\end{figure}

Consider a group of $N=4$ agents with the dynamics described by (\ref{eq:agent})
and the system matrices given as follows
\begin{gather*}
A_{i}(w_{i})=\left[\begin{array}{cccc}
-1+w_{i,1} & 1 & 0 & 0\\
0 & 0 & 1 & 1\\
1 & -2+w_{i,2} & -2+w_{i,3} & 0\\
3 & 0 & 1 & 2
\end{array}\right], \;
B_{i}(w_{i})=\left[\begin{array}{cc}
0 & 0\\
0 & 1\\
2+w_{i,4} & 0\\
0 & 0
\end{array}\right],\;C_{i}(w_{i})=\left[\begin{array}{cccc}
1 & 0 & 0 & 0\\
0 & 0 & 0 & 1
\end{array}\right],
\end{gather*}
All the uncertainties $w_{i,j}$ for $i=1,\cdots,N$ and $j=1,\cdots,4$
vary within $[-1,1]$. The objective is to synchronize all the agents\textquoteright{}
outputs $y_{i}$ in a pattern described in (\ref{eq:pattern}) with
\[
A_{o}=\left[\begin{array}{cc}
0 & 0.5\\
-0.5 & 0
\end{array}\right],\;C_{o}=\left[\begin{array}{cc}
1 & 0\\
0 & 1
\end{array}\right].
\]
The communication network of multi-agent system is illustrated in
Fig. \ref{fig:net} with Laplacian matrix
\[
\mathcal{L}=\left[\begin{array}{cccc}
3 & -2 & -1 & 0\\
-1 & 2 & 0 & -1\\
0 & 0 & 1 & -1\\
0 & 0 & -1 & 1
\end{array}\right].
\]

We first design reference model (\ref{eq:RM}) for each agent with
\begin{gather*}
B_{o}=\left[\begin{array}{c}
0\\
35
\end{array}\right],\;A_{ \zeta}=\left[\begin{array}{cc}
-0.03 & 0.47\\
-0.357 & -3.94
\end{array}\right]\\
B_{ \zeta}=0.03\left[\begin{array}{cc}
1 & 1\\
1 & 1
\end{array}\right],\;C_{ \zeta}=\left[\begin{array}{cc}
0.087 & 0.112\end{array}\right]
\end{gather*}
which is confirmed to be able to achieve perturbed consensus problem with $\gamma_\zeta =0.139$.

Then, we can get the solution to the regulator equation (\ref{eq:RE})
with $w_{i}=0$ as follows
\[
X_{i}= \left[\begin{array}{cccc}
1 & 1    & -3.5 & 0\\
0 & 0.5 & -2 & 1
\end{array}\right]\t,
\; U_{i}=\left[\begin{array}{cc}
-2.5 & 2.375\\
3.25 & 1.5
\end{array}\right]
\]
which can be used to calculate $\Upsilon_{i}$.
We can calculate
\[
\begin{array}{cc}
\Phi_{i}=I_{2}\otimes \left[\begin{array}{cc}
0 & 1 \\
-0.25 & 0
\end{array}\right],& \;\Psi_{i}=I_{2}\otimes \left[\begin{array}{cc}
1 & 0 \end{array}\right].\end{array}
\]
The internal model
(\ref{eq:IM}) is designed with
\[
\begin{array}{cc}
M_{i}=I_{2}\otimes\mbox{diag}\{-0.5,-1\},\; & N_{i}=I_{2}\otimes \left[\begin{array}{cc}
1 & 1 \end{array}\right]\t\end{array}.
\]
The solution to the Sylvester equation (\ref{syl}) gives $T_i$.
According to Lemma~\ref{lem:mas:stb}, we design the controller
(\ref{eq:mas-stb}) with
\[
K_{i}=\left[\begin{array}{cccc}
112 & 0 & 40 & 787\\
1188 & 81 & 0 & 137
\end{array}\right].
\]
and $L_i$ such that $\bar{A}_i-L_i\bar{C}_i$ Hurwitz,
which can achieve the $\gamma$-stabilization problem for the system
(\ref{eq:Chi}) with $\gamma<1/(N\gamma_\zeta) =1.8$.

\begin{figure}
\centering\includegraphics[scale=0.4]{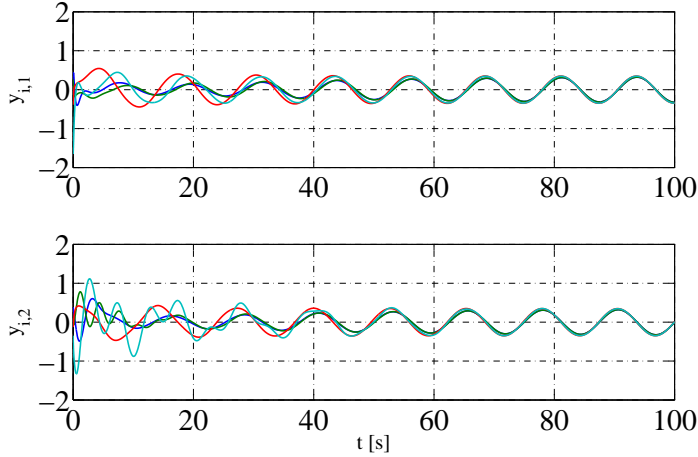}\protect\caption{The synchronization of two output components $y_{i,1}$ and $y_{i,2}$ \label{fig:syn}}

\end{figure}

By Theorem~\ref{thm:outputSyn}, the robust output synchronization problem
is solved by a distributed controller of the form (\ref{mas-overallcontroller})
with all the parameters explicitly calculated above.  The performance of output synchronization
is illustrated in Fig.\ref{fig:syn} where all the agent outputs converge to the common sinusoidal
waveform whose frequency is $0.5$~rad/s determined by $A_o$ and amplitudes and phases by
the initial values of the closed-loop MAS.

\section{Conclusion\label{sec:con}}

In the first part of the paper, we have identified a class of MIMO systems for which
the $\gamma$-stabilization problem can be solved.
In particular, we  have proposed a static state feedback controller for
the system in the normal form and a  dynamic output feedback controller for
a class of systems with a particular structure that arises from the
robust output synchronization problem of MASs.
In the second part of the paper,
we  have proposed a modified
internal model design such that the robust output synchronization
problem can be converted into perturbed consensus and perturbed
regulation problems. The latter can be further converted to the $\gamma$-stabilization problem
which can be solved by the dynamic output feedback controller developed in the first part.
Overall, the robust output synchronization problem is solved by a distributed output feedback/communication
controller integrating the techniques of
reference model, internal model and $\gamma$-stabilization.

\section{Appendix}
\textbf{Proof of Lemma \ref{lem:AB}:}
First, we note that
 $(\Phi_{0},\Gamma_{0})$
is controllable and $\Gamma_0\neq 0$ ($r_0>0$) by Assumption~\ref{ass:con}.
We will prove the lemma using mathematical induction.
Suppose, for $j\geq 0$, $(\Phi_{k},\Gamma_{k})$
is controllable  and $\Gamma_k \neq 0$ (i.e., $r_k > 0$) for $k=0,\cdots, j$.
If $\Gamma_j$ has a full row rank, the lemma is true. Otherwise,
the algorithm gives a valid pair of $(\Phi_{j+1},\Gamma_{j+1})$.
Next, we aim to show that $(\Phi_{j+1},\Gamma_{j+1})$
is controllable  and $\Gamma_{j+1} \neq 0$ (i.e., $r_{j+1} >0$).

Denote
\EQQ
Q(\lambda)=\left[\begin{array}{cc}
\lambda I-\Phi_{j} & \Gamma_{j}\end{array}\right].
\ENN
By PBH test, the fact that $(\Phi_{j},\Gamma_{j})$ is controllable
implies that
\EQQ \mbox{rank}(Q(\lambda))=n-\Sigma_{k=0}^{j-1}r_{k},\;
\forall  \lambda\in\mathbb{C}.
\ENN
Let
\EQQ
\bar{Q}(\lambda)=U_{j+1}\t Q(\lambda) \left[\begin{array}{cc} U_{j+1} & \\ & H_{j+1} \end{array}\right].
\ENN
One has $\mbox{rank}(\bar Q(\lambda)) = \mbox{rank}(Q(\lambda))$
as  $U_{j+1}$ and  $H_{j+1}$ are
unitary matrices.
Direct calculation shows that
\EQQ
\bar{Q}(\lambda)=\left[\begin{array}{cccc}
\lambda I-\tilde{U}_{j+1}\t\Phi_{j}\tilde{U}_{j+1} & -\tilde{U}_{j+1}\t\Phi_{j}\bar{U}_{j+1} & \Sigma_{j} & 0\\
-\bar{U}_{j+1}\t\Phi_{j}\tilde{U}_{j+1} & \lambda I-\bar{U}_{j+1}\t\Phi_{j}\bar{U}_{j+1} & 0 & 0
\end{array}\right]\\
=\left[\begin{array}{cccc}
\lambda I-\tilde{U}_{j+1}\t\Phi_{j}\tilde{U}_{j+1} & -\tilde{U}_{j+1}\t\Phi_{j}\bar{U}_{j+1} & \Sigma_{j} & 0\\
-\Gamma_{j+1} & \lambda I- \Phi_{j+1} & 0 & 0
\end{array}\right].
\ENN
Since $\Sigma_{j}$ has a full rank, i.e.,  $\mbox{rank}(\Sigma_{j}) = r_j$, one has
\begin{equation}
\mbox{rank}\left(\left[\begin{array}{cc}
\lambda I-\Phi_{j+1} & \Gamma_{j+1}\end{array}\right]\right)=
\mbox{rank}(\bar Q(\lambda)) -r_j =
n-\Sigma_{k=0}^{j}r_{k},\; \forall  \lambda\in\mathbb{C}. \label{eq:AjBj}
\end{equation}
Applying PBH test again shows that $(\Phi_{j+1},\Gamma_{j+1})$ is controllable
and $\Gamma_{j+1} \neq 0$ (i.e., $r_{j+1} >0$).

From mathematical induction,  for all $j\geq 0$, the pair $(\Phi_{j},\Gamma_{j})$
is well defined by the algorithm and controllable,
until there is a finite number $l$ such that $\Gamma_{l}$ has a full row rank.
Such a finite number $l$ always exists because the square matrix $\Phi_j$
of the dimension  $n-\Sigma_{k=0}^{j-1}r_{k}$ with $r_k >0$ cannot be well defined
for an arbitrarily large $j$.  \eproof

\textbf{Proof of Lemma \ref{lem:y}}: Let
\EQQ
\bar{C}_{0}=C,\;
\bar{C}_{j}=\bar{C}_{j-1}\bar{U}_{j},\; C_{j}=\bar{C}_{j-1}\tilde{U}_{j},\;j=1,\cdots,l.
\ENN
Next, we will use mathematical induction to prove, for $j=1,\cdots,l$,
\EQ
{\rm Claim-}j: \;\;
\left\{
\begin{split}
& \bar{C}_{k}\Phi_{k}^{j-1-k}\Gamma_{k}  = 0,\;k=0,\cdots,j-1  \\
& C_{j}  =  0  \label{eq:CAB_ind}
\end{split}\right. .
\EN

For $j=1$, one has
\EQQ
CB= C\Gamma_0 =C\tilde{U}_{1}\left[\begin{array}{cc}\Sigma_{0} & 0 \end{array}
\right]H_{1}\t=0
\ENN
by (\ref{eq:svd_Bj}) and Assumption~\ref{ass:level}. Since both $\Sigma_{0}$
and $H_1$ are nonsingular, the above equations imply $C_{1}=C\tilde{U}_{1}=0$. Thus,
Claim-$1$ is proved.

For $1 \leq j \leq l-1$, we assume Claims-($1, \cdots, j$) hold
and will prove Claim-$(j+1)$.
Claims-($1, \cdots, j$) mean that
\EQQ
\left\{
\begin{split}
& \bar{C}_{k}\Phi_{k}^{s-1-k}\Gamma_{k}  = 0,\;k=0,\cdots,s-1  \\
& C_{s}  =  0
\end{split}\right. ,\;\;  s=1,\cdots,j.
\ENN
that, with re-organization, implies
\EQ
\left\{
\begin{split}
&\bar{C}_{k}\Phi_{k}^{s-1}\Gamma_{k}  =  0,\;s=1,\cdots,j-k \\
&C_{k} =  0,\;
\end{split} \right. ,\; \; 1 \leq k \leq j . \label{eq:CAB_ind-1}
\EN

With (\ref{eq:CAB_ind-1}) in hand, we can prove the following implication
\EQ
\bar{C}_{k-1}\Phi_{k-1}^{j-(k-1)}\Gamma_{k-1}=0 \; \Longrightarrow \;\bar{C}_{k}\Phi_{k}^{j-k}\Gamma_{k}=0,\;
1 \leq k \leq j . \label{implication}
\EN
To prove (\ref{implication}), we denote
\begin{equation}
\Delta=\left[\begin{array}{c}
\begin{array}{c}
\tilde{U}_{k}\t\\
\bar{U}_{k}\t
\end{array}\end{array}\right]\Phi_{k-1}\left[\begin{array}{cc}
\tilde{U}_{k} & \bar{U}_{k}\end{array}\right]=\left[\begin{array}{cc}
* & * \\
\Gamma_{k} & \Phi_{k}
\end{array}\right]. \label{eq:Aj_t}
\end{equation}
Note that
\EQQ
\bar{C}_{k-1}\Phi_{k-1}^{j-(k-1)}\Gamma_{k-1} & = & \bar{C}_{k-1}\left[\begin{array}{cc}
\tilde{U}_{k} & \bar{U}_{k}\end{array}\right]
\Delta^{j-(k-1)}\left[\begin{array}{c}
\begin{array}{c}
\tilde{U}_{k}\t\\
\bar{U}_{k}\t
\end{array}\end{array}\right]\left[\begin{array}{cc}
\tilde{U}_{k} & \bar{U}_{k}\end{array}\right]\left[\begin{array}{cc}
\Sigma_{k-1} & 0\\
0 & 0
\end{array}\right]H_{k}\t\\
 & = & \left[\begin{array}{cc}
0 & \bar{C}_{k}\end{array}\right]\Delta^{j-(k-1)}\left[\begin{array}{cc}
\Sigma_{k-1} & 0\\
0 & 0
\end{array}\right]H_{k}\t \ENN
where $C_{k}=\bar{C}_{k-1}\tilde{U}_{k}=0$
is used. Substituting (\ref{eq:Aj_t}) to the above equation and
using the first equation of (\ref{eq:CAB_ind-1}) lead to
\EQQ
\bar{C}_{k-1}\Phi_{k-1}^{j-(k-1)}\Gamma_{k-1}=
\bar{C}_{k}\Phi_{k}^{j-k}\Gamma_{k}
\left[\begin{array}{cc}\Sigma_{k-1} & 0 \end{array}
\right]
H_{k}\t=0.
\ENN
So, one has  $\bar{C}_{k}\Phi_{k}^{j-k}\Gamma_{k}=0$ as both
$\Sigma_{k-1}$ and $H_{k}$ are nonsingular. The proof  of  (\ref{implication}) is thus complete.

By recursively using (\ref{implication}),
Assumption~\ref{ass:level}, i.e.,
$CA^{j}B=\bar{C}_{0}\Phi_{0}^{j}\Gamma_{0}=0$
implies $\bar{C}_{1}\Phi_{1}^{j-1}\Gamma_{1}=0$,
$\bar{C}_{2}\Phi_{2}^{j-2}\Gamma_{2}=0$, until
$\bar{C}_{j}\Phi_{j}^{0}\Gamma_{j}=0$.
In summary,  $\bar{C}_{k}\Phi_{k}^{j-k}\Gamma_{k}=0,\;k=0,\cdots,j$.

Next, we note from (\ref{eq:svd_Bj}) that
\begin{eqnarray*}
\bar{C}_{j}\Gamma_{j} & = & \bar{C}_{j}\left[\begin{array}{cc}
\tilde{U}_{j+1} & \bar{U}_{j+1}\end{array}\right]\left[\begin{array}{cc}
\Sigma_{j}&0\\
0&0
\end{array}\right]H_{j+1}\t\\
 & = & C_{j+1}
 \left[\begin{array}{cc}\Sigma_{j} & 0 \end{array}
\right]
 H_{j+1}\t=0
\end{eqnarray*}
which implies $C_{j+1}=0$ as both
$\Sigma_{j}$ and $H_{j+1}$ are nonsingular.
From above, we have proved Claim-$(j+1)$.

\medskip

For $j=2,\cdots, l+1$, one has
$C T_j\t = C_{l-j+2}=0$ by Claim-$(l-j+2)$. Then
\EQQ
y=Cx=CT\t  \zeta=
C T_1\t  \zeta_{1}
+\sum_{j=2}^{l+1} CT_j\t  \zeta_j
=C T_1\t  \zeta_{1}.
\label{eq:y_xi}
\ENN
The proof is thus completed.
 \eproof

\bibliographystyle{unsrt}
\bibliography{literature}

\end{document}